**Frontiers of Condensed Matter Physics Explored with High-Field Specific Heat**

Marcelo Jaime

MPA-CMMS, Los Alamos National Laboratory, Los Alamos, NM 87545, USA.

**Abstract**

Production of very high magnetic fields in the laboratory has relentlessly increased in quantity and quality over the last five decades, and a shift occurred from research focused in magnet technology to studies of the fundamental physics of novel materials in high fields. New strategies designed to understand microscopic mechanisms at play in materials surfaced, with methods to extract fundamental energy scales and thermodynamic properties from thermal probes up to 60 tesla. Here we summarize developments in the area of specific heat of materials in high magnetic fields, with focus in the original study of the Kondo Insulator system $Ce_3Bi_4Pt_3$.

## 1. Introduction

One of the most remarkable features of the present times, from the standpoint of the balance between basic science and technology applications, is that while magnetic materials and magnetism have central roles in a large part of our life, ubiquitous to motor generators, transportation, electronic devices, medical diagnostics tools, and industrial processes of many sorts, we have yet to fully understand the origin of magnetism and the consequences of a strong interplay between magnetic, charge and lattice degrees of freedom in materials. High magnetic fields are instrumental to tackle this challenge. Indeed, the effect of high magnetic fields on materials includes favoring exotic magnetic states in detriment of non-magnetic ones, inducing changes in electronic band structure through manipulation of spin-orbit coupling and magnetostriction, suppression of superconductivity, inducing metal-insulator phase transitions, driving charge and spin excitations through dimensional crossovers and quantum limits, tilting the balance between competing ground-states, and tuning quantum fluctuations near quantum critical points, among a number of other effects. A handful of experimental techniques have historically been the tools-of-choice for the study of these phenomena, *i.e.* mostly conventional and quantum Hall effect, magnetoresistance, magnetization, ESR/EPR, de Haas-van Alphen effect, Suvnikov-de Haas effect, and NMR. However, thermal properties remained largely out of the radar screen of experimentalists due to some significant technical obstacles, such as reliable thermometry in high fields and fast varying magnetic fields during experiments. In this work we summarize our strategy to overcome technical obstacles and achieve some degree of maturity in the field of measurement of thermal properties and specific heat in magnetic fields to 60 tesla. We have done so by removing some low-temperature-physics experimental taboos, and by shedding light on poorly understood thermal processes. During this quest we benefited from the talent of numerous material scientists that produced samples of extraordinary quality and fascinating physics. In many of the case studies, a significant theoretical effort served as the intellectual driving force for progress, with solid interpretations and encouraging predictions.

## 2. Specific heat at high magnetic fields: a brief recount

Specific heat at constant pressure ($C_p$) measurements in elements and compounds has been used for more than half a century to understand their properties, approximately since the time when Brown, Zemansky and Boorse[1] first measured the low temperature $C_p(T)$ of the superconducting and field-induced normal states of pure Niobium down to T = 2K, in a magnetic field of 6000 gauss. This development was, not surprisingly, tied to the serendipitous discovery of negligible magnetoresistance in what would become the thermometry of choice for decades to come: radio carbon resistors (also known as *carbon-glass*) manufactured by Allen-Bradley Co. While there are probably thousands of publications of specific heat in the presence of man-made magnetic fields, a comprehensive discussion of all contributions is both impossible and out of the scope of this work. Having said so, a quick bibliographic search indicates that by 1958 standard magnetic fields available increased to 10,000 gauss[2], and soon were cranked up to 70,000 gauss to study the robust superconducting state in $V_3Ga$, when Morin et al. estimated that fields as high as 300,000 gauss (30 tesla) were necessary to completely suppress superconductivity in $V_3Ga$[3], a development that would take several decades. As a matter of fact, as late as 1980 standard laboratory fields achieved with superconducting coils remained close to 10 tesla, as discussed for instance by Ikeda and Gschneidner[4] among others. By these times the calorimetry methods of choice were the adiabatic method[2] or heat-pulse[5] method with relatively large samples (several grams), and thermometry had migrated to lightly doped germanium. The development of AC calorimetry[6] and the thermal relaxation time technique[7] made calorimetry a real option for small high-quality single crystalline samples in the mid 70's, allowing for the first measurements up to 18 tesla[8], but optimal thermometry was still limited. The development of much smaller $RuO_2$ resistive-paste thermometry and 24-26 tesla hybrid resistive/superconducting magnets opened the window to the first specific heat and magnetocaloric effect studies of the metamagnetic phase transition in $UPt_3$[9], that shows a critical field $\mu_0 H_c$ = 20.3 tesla. With the commissioning of the 60 tesla long pulse magnet (see Fig 2 Top) at the National High Magnetic Field Laboratory (NHMFL) in 1998 the evolution of magnetic fields suitable for specific heat experiments witnessed a significant step up with the measurements of the Kondo insulator $Ce_3Bi_4Pt_3$[10], using a quasi-adiabatic technique in a setup furbished with calibrated bare-chip Cernox® thermometry (Fig 2 Bottom), a record that still holds unchallenged. The construction and operation of the world class 45 tesla hybrid magnet at the NHMFL was, especially for calorimetrists, truly revolutionary. Indeed, since the first measurements that elucidated a evasive (H,T) phase diagram in $URu_2Si_2$ done in 2002[11], approximately more than a dozen different specific heat experiments were run by several groups. Closer to our interest are a number of studies of quantum magnetism, heavy fermions and superconductors[12,13] that, using various technique developments[14,15] topped this year with some re-exploration of AC-$C_p$ and magnetocaloric effect in pulsed field magnets[16]. The next technical frontier for specific heat in high magnetic fields is the full development of a method suitable for 20 msec long pulses that can reach the 100 tesla territory, but it will require significant improvements over presently known techniques.

## 3. Specific heat in pulsed magnetic fields

There are enough technical challenges in the field of pulsed magnets to initially discourage most condensed matter physics experimentalists, and number one in the list of obstacles is the short duration of magnetic field pulses. Following in importance is the generalized perception that Eddy-current heating induced by rapidly changing magnetic fields in the samples under study inevitably leads to poor data quality. Some other popular reasons for aversion to pulsed-fields include electromagnetic noise, thermometry, mechanical vibrations, reduced sample space, magnetocaloric effect (experiments are usually adiabatic, not isothermal like in DC magnetic fields), lack of equilibrium between sample and magnetic fields and, last but not least, the concerns related to catastrophic failure of the magnets and cryogenic equipment with the concomitant irreversible loss of the specimens. In what follows we describe how these obstacles are solved or circumvent in the case of specific heat measurements.

Dealing with short duration magnetic field pulses in calorimetry requires several considerations, regarding principally thermal equilibrium between sample and thermometry, and measurement techniques for resistive temperature sensors. The first point, thermal equilibrium within the duration of the field pulse, is addressed with the miniaturization of the specific heat stage. In our case we used a 6x6x0.25 mm$^3$ platform Si (single crystal) platform, onto which we glued a 1x1x0.25 mm$^3$ heater made of amorphous-metal film on a Si substrate, a 1x0.5x0.25 mm$^3$ bare chip Cernox® thermometer, and the sample under study. To minimize Eddy-current heating during magnetic field pulses we mounted all elements parallel to the applied magnetic field, and used 25 micrometer resistive-alloy electrical wires for connections. The samples, weighting typically 1-50 mg depending on availability and expected contribution to the total specific heat, were polished into slabs as thin as the material would allow (typically 200-400 micrometers) with the largest surface glued to the Si platform to minimize the internal thermal time constant in the calorimeter platform, and also to minimize the sample cross section in the direction of the applied field. Extensive experience with state-of-the-art high frequency AC techniques in our lab indicates that measurement of electrical resistances between ~1 ohm and ~$10^3$ ohm are relatively easy to accomplish in a sub-millisecond timeframe, and that resistances between ~$10^{-3}$ ohm and ~$10^5$ ohm are detectable, with anything smaller or larger requiring somewhat extraordinary measures and highly skilled experimentalists.

The first calorimeter used for experiment in pulsed fields to 60 tesla is displayed in Fig 3. The picture shows a non-metallic frame made of epoxy embedded fiberglass (G-10), and used to attach the nylon strings holding a Si platform. On the top of the platform it is possible to see the resistive film heater, the opposite side holds the bare-chip thermometer. Electrical connections are made with 5 inches-long NiCo alloy (Constantan®) minimizing open electrical loops. The copper coil at the far left end is used to locally measure the voltage induced by the time-varying magnetic field during the pulse. The main thermometer, an un-encapsulated Cernox® resistor is located at the far right and glued directly on the Si block. This setup sits in vacuum during the experiment, and the vacuum can was made of StyCast® 1266 epoxy resin manufactured in-house with a conical seal.

Essential to the proper functioning of any calorimeter in high magnetic fields is the availability of calibrated thermometers. In our case we choose a bare chip CX-1030 Cernox® thermometer having a resistance of approximately 900 ohm at 4 kelvin. The calibration of this chip was done in a 25 msec 60 tesla pulsed magnet, immersed in liquid/gas $^4$He, with the magnetic field applied in the plane of the resistive film, perpendicular to contact pads, *i.e.* perpendicular to the applied AC current (see Fig 3 bottom panel). We choose the quasi-adiabatic method to measure the specific heat in pulsed fields with our setup. In this method a pulse of current is delivered to the heater on the specific heat platform, and the temperature of the entire ensemble (platform + sample + thermometer + heater) is recorded as a function of time.

## 4. Specific heat of $Ce_3Bi_4Pt_3$ at high magnetic fields

Kondo insulator materials (KIM), such as $Ce_3Bi_4Pt_3$, are intermetallic compounds that behave like simple metals at room temperature but an energy gap opens in the conduction band at the Fermi energy when the temperature is reduced[17]. The formation of the gap in KIM is not yet completely understood, and has been proposed to be a consequence of hybridization between the conduction band and the *f*-electron levels in Ce[18]. We used an external magnetic field (H) to close the charge/spin gap in $Ce_3Bi_4Pt_3$, and observed a significant increase in the Sommerfeld coefficient γ(H), consistent with a magnetic field-induced metallic state. Indeed, here we present specific-heat measurements of $Ce_3Bi_4Pt_3$ in pulsed magnetic fields up to 60 tesla. Numerical results and the analysis of our data using the Coqblin-Schrieffer model demonstrated unambiguously a field-induced insulator-to-metal transition[10].

The raw unprocessed data collected during the experiment using a 44.85 mg single crystal sample are displayed in Fig 4, where the color-coded plots show the temperature of the calorimeter (black), the magnetic field measured with the Cu-coil near to the sample (red), and voltage across the sample heater (blue) vs time. Data for four different pulses are plotted in Fig 4A, where two specific heat data points are obtained per pulse. The specific heat is calculated in the calorimeter simply as the energy in joules delivered by the heater, calculated as $E = \int V(t).I(t)dt$, where $V(t)$ and $I(t)$ are the time dependent voltage and current in the heater respectively, divided by the temperature change $\Delta T$. Fig 4B shows eight specific heat data points collected in a single 41.5 tesla pulse, when the voltage in the heater is programmed conveniently. A compilation of data obtained with the calorimeter is shown in Fig 5. In the top panel the specific heat of $Ce_3Bi_4Pt_3$ divided by the temperature ($C_p/T$) is plotted vs temperature square, after subtraction of the empty calorimeter (addenda) contribution. The bottom panel contains data for a 300 mg Silicon sample, without subtracting the addenda. The simplest expression for the specific heat of a metal is $C_p/T = \gamma + \beta T^2$, where $\gamma$ is the Sommerfeld coefficient due to free electrons, $\beta$ is the lattice contribution, and T is the temperature. While in the case of Silicon neither the slope ($\beta$) nor the small T$\rightarrow$0 extrapolation are affected by a field of 60 tesla within the experimental scattering of the data, in the case of $Ce_3Bi_4Pt_3$ the electronic contribution $\gamma(H)$ is a strong function of the magnetic field.

The sample measured here shows a finite $\gamma(H=0) = 18.6$ mJ/molK$^2$ that is roughly 2/3 of the value observed in the metallic system $La_3Bi_4Pt_3$. This value, high for an insulator, has an unclear origin. While extrinsic effects such as impurities, vacancies and CeO surface states can certainly contribute, a recent proposal suggests that in-gap states could be magnetic-exciton bound states intrinsic to KIM[19,20]. Magnetocaloric effect (MCE) results obtained during this experiment can provide additional evidence to this effect. Fig 6 Top shows the extracted $\gamma$ vs H for $Ce_3Bi_4Pt_3$, and the bottom panel displays the MCE traces recorded during the experiment.

In order to tell whether our results indicate the recovery of the metallic Kondo state in high fields, the increase in $\gamma(H)$ observed in $Ce_3Bi_4Pt_3$ needs to be put in perspective. Using an estimate for the Kondo temperature of $T_K^0 = 240\text{-}320$ K, the Sommerfeld coefficient for a metal with such $T_K$ can be estimated using the expression for a single-impurity Kondo[21] to be K = 53 - 70 mJmol$^{-1}$K$^{-2}$. This provides an upper bound for the high field $\gamma(H)$ in Ce$_3$Bi$_4$Pt$_3$, as it is expected that an external field will suppress correlations and induce a reduction in $\gamma(H)$. Taking into account the effect of the applied magnetic fields within a single impurity model[22], our estimate of the Sommerfeld coefficient at 60T is $\gamma(60T) = 51\text{-}66$mJ mol$^{-1}$K$^{-2}$. Hundley et al.[23] have measured the compound La$_3$Bi$_4$Pt$_3$ in zero field, and obtained $\gamma_{La} = 27$mJmol$^{-1}$K$^{-2}$. This value in La$_3$Bi$_4$Pt$_3$, an isostructural metal where electronic correlations are absent, should be our lower bound limit on $\gamma(60T)$ in the high field metallized state of Ce$_3$Bi$_4$Pt$_3$.

The upper panel in Fig 6 shows $\gamma(H)$ for Ce$_3$Bi$_4$Pt$_3$ in applied magnetic fields up to 60 T. The values were obtained from a single-parameter fit of the form $C(T) = \gamma T + \beta T^3$, with the coefficient $\beta$ of the lattice term fixed to its zero-field value. We see a sharp rise in $\gamma(H)$ between H = 30T and 40T. The result of the fit suggests a saturation at a value of $\gamma^{sat}(H) = 62\text{-}63$mJmol$^{-1}$K$^{-2}$ above 40 T. The strong enhancement of $\gamma(H)$ from its zero-field value, and the quantitative agreement with the estimate based on $T_K$ for a metallic ground state of Ce$_3$Bi$_4$Pt$_3$, prove that we indeed crossed the phase boundary between the Kondo insulator and the Kondo metal.

Additional evidence for a significant change of regime induced by magnetic fields in Ce$_3$Bi$_4$Pt$_3$ is found in the magnetocaloric effect, *i.e* the changes in sample temperature when the magnetic field is swept in adiabatic conditions. The lower panel of Fig 6 shows the temperature vs magnetic field (same as T vs time curves in Fig 4) for all magnetic fields used to determine $C_p(T,H)$. The curves show an initial increase (①) that could be due to alignment of paramagnetic impurities, then a decrease of the temperature (②) consistent with the increase in $\gamma(H)$ displayed in the top panel. At top field (③) the temperature increases due to the heat pulses delivered to the platform. Finally, when the magnetic field decreases the temperature of the sample reversibly increases (④) as $\gamma(H)$ drops to the low field value, to then decrease in the impurity zone (⑤). One important observation from this drop is that the initial (low field) temperature change in T(H) decreases in magnitude as the initial sample temperature rises from ~1.8 K to ~4.5 K. On the other hand the temperature change observed in the high field region, where $\gamma(H)$ is a strong function of H, is roughly temperature independent (parallel MCE curves). A quick analysis of the MCE effect curves reveals that the low field region ($0 < \mu_0 H < 16T$) has a much stronger temperature dependence than the high field region ($16T < \mu_0 H < 60T$). Fig 7 displays the absolute temperature change $|\Delta T|$ extracted from the curves in Fig 6 lower panel. The high field region (red circles), clearly linked to $\gamma(H)$ and hence linked to the charge gap, depends only slightly on temperature likely due to relative changes between electronic and phononic contributions as the temperature is increased. Indeed as the phonon contribution increases the entropy change observed upon closing of the charge gap, which remains the same, causes a gradually smaller $|\Delta T|$. On the other hand, the temperature change observed at low fields (blue triangles) changes much more rapidly with temperature, in a region of magnetic field where $\gamma(H)$ changes little. The rapid change resembles the temperature dependence of the magnetic susceptibility and leads us to conclude that in this region $|\Delta T|$ is dominated by free magnetic moments. Indeed, paramagnetic magnetic moments in external fields produce positive MCE. The origin of the magnetism is unclear, but we feel that impurities alone cannot explain the magnitude of the observed effect, and that some type of intrinsic phenomenon must play an important role[20].

## 5. Determination of Phase Diagrams in High Fields

Since the first measurements of specific heat in the 60 tesla long pulse magnet at NHMFL-Los Alamos many interesting materials surfaced that have strong correlations between charge, spin and lattice degrees of freedom. High magnetic fields can be used to tilt the balance between competing mechanisms, just like in the case of $Ce_3Bi_4Pt_3$, to study phases of matter that do not otherwise occur. Fig 8 shows a compilation of data for all materials and systems where strong magnetic fields were used to change the ground state[12]. Figure 8 Top displays phase diagrams for strongly correlated metals, including the superconductor $La_2CuO_{4.11}$ where H//c-axis, the valence transition material $YbInCu_4$, the hidden order-parameter compound $URu_2Si_2$, and AFM metals $CeIn_3$ and $CeIn_{2.75}Sn_{0.25}$. Note that all these materials present a low temperature state (different in nature) that can be suppressed with a strong enough magnetic field, and that the energy scale of the zero-field phase transition ($T_c$, $T_V$, $T_N$) does not correlate in a simple manner with the magnetic field required to change the ground state. The reason for this is, clearly, the diversity of zero-field ground states observed and the spread in magnitude in the coupling between order parameter and magnetic field[12]. $Ce_3Bi_4Pt_3$, which does not show phase boundaries but a crossover, is not displayed here.

Fig 8 Bottom displays (T,H) phase diagrams for a collection of classical and quantum magnets (insulators). The magnetic system $RbFe(MoO_4)_2$ presents a in-plane AFM phase in zero field that turns into a multiferroic phase as a modest magnetic field of a few tesla is applied. The rest of the materials displayed do not show magnetism in zero field, but a XY-type AFM state is induced by the applied magnetic field. The strength of the exchange interactions, the always present geometrical frustration, and the magnetic lattice dimensionality play together, or against each other, to determine the magnitude of magnetic fields necessary to induce a change of ground state. These materials, also known as quantum magnets, can under special circumstances be approximated with a model that describes Bose-Einstein condensation of magnons [12,24]. The insulator-to-metal crossover in $Ce_3Bi_4Pt_3$ is displayed for comparison purposes.

Specific heat studies in high magnetic field have been instrumental to understand the mechanisms and physics at play in correlated electron systems and quantum magnets, and we do expect to continue finding new materials that will help us understand many physics puzzles that still exist in these topical areas. Some of these outstanding puzzles include the nature of magnetic states in $Ce_3Bi_4Pt_3$, the pairing mechanisms in cuprate superconductors, the nature of the order parameter in $URu_2Si_2$, and the effects of geometrical frustration, quantum fluctuations and lattice coupling in quantum magnets. In particular, one area of development for the near future will be the implementation of AC specific heat to determine the shape of the phase boundary and relevant physics of quantum magnets in the high magnetic field region, *i.e.* for fields $\mu_0 H > 45T$ and temperatures below 4 kelvin.

**Acknowledgements:** Work at the National High Magnetic Field Laboratory was supported by the US National Science Foundation, the State of Florida and the US Department of Energy through Los Alamos National Laboratory.

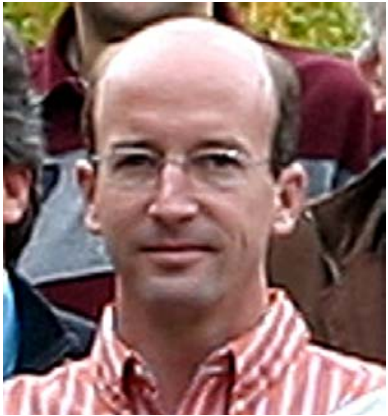

Marcelo Jaime
MPA-CMMS, Los Alamos National Laboratory, Los Alamos, NM 87545, USA.
TEL. +1-505-667-7625
FAX. +1-505-665-4311
e-mail: mjaime@lanl.gov
Research area: Specific heat of materials at very high magnetic fields and low temperatures.

**Figure Captions**

**Fig. 1** A selection of representative specific heat experiments performed in man-made magnetic fields, plotted as strength of magnetic field vs year of publication. Bracketed numbers indicate references.

**Fig. 2 Top:** from right to left D. Rickel, J. Schillig, R. Movshovich and the author standing in front of the first motor generator-driven 60 tesla Long Pulse Magnet, equipped with a $^4$He cryostat. **Bottom:** magnetic field pulses. The purple line in the center is a pulse produced by a standard capacitor bank-driven pulsed magnet.

**Fig. 3 Top:** first functional calorimeter for pulsed magnetic fields[10], consisting of a frame made of G10 attached to a Si block. **Bottom:** electrical resistance vs magnetic field for the Cernox® bare chip thermometer.

**Fig 4.** Color coded plot of sample Temperature (black), applied Magnetic field (red) and sample Heater voltage (blue) vs time recorded during a $C_p$ experiment in the 60 tesla Long Pulse Magnet. **A)** $C_p$ measurement in 42 a tesla pulse **B)** $C_p$ measurements in 60 tesla pulses.

**Fig. 5 Top:** specific heat divided by the temperature $C_p/T$ vs $T^2$ for $C_3Bi_4Pt_3$ for magnetic fields up to 60T as indicated in the label. A sudden change is observed between H = 30 tesla and 40 tesla. **Bottom:** $C_p$ (T,H) results for Silicon in H = 0 and H = 60T

**Fig. 6**. Top: Sommerfeld coefficient γ vs. field for $Ce_3Bi_4Pt_3$. The value observed in $La_3Bi_4Pt_3$ is indicated as comparison. **Bottom:** temperature of the specific heat platform + sample vs field, as recorded during the specific heat experiment.

**Fig. 7** Absolute value of the temperature change observed in $Ce_3Bi_4Pt_3$ when the magnetic field is swept. Blue triangles correspond to the low field region. Red circles, were extracted at high fields.

**Fig. 8 Top:** (T,H) phase diagrams determined from $C_p(T)$ for $La_2CuO_{4.11}$, $YbInCu_4$, $URu_2Si_2$, $CeIn_3$ and $CeIn_{2.75}Sn_{0.25}$. **Bottom:** (T,H) phase diagrams determined from $C_p(T,H)$ and MCE experiments for $RbFe(MoO_4)_2$, $NiCl_2$-$4SC(NH_2)_2$, $Ba_3Mn_2O_8$, $Ba_3Cr_2O_8$, $BaCuSi_2O_6$ and $Sr_3Cr_2O_8$. Star symbols (☆) from magnetization vs field. For more details see ref. (12)

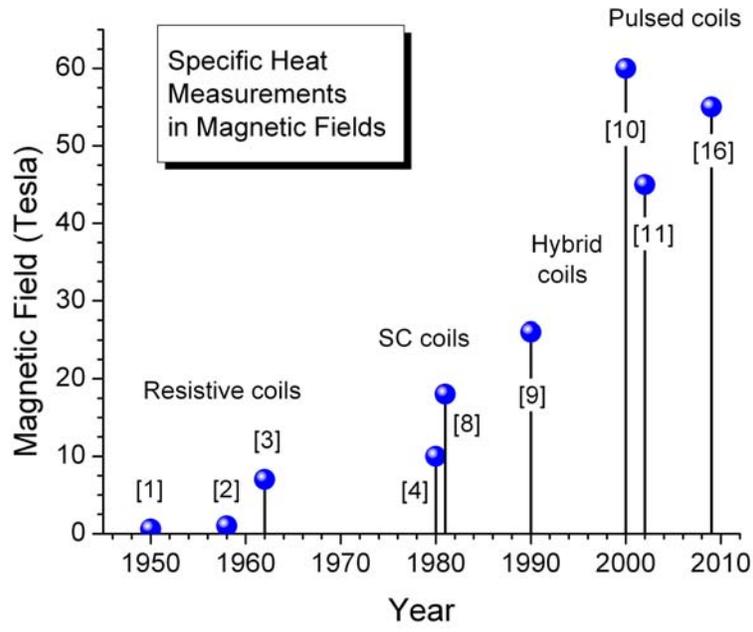

Figure 1

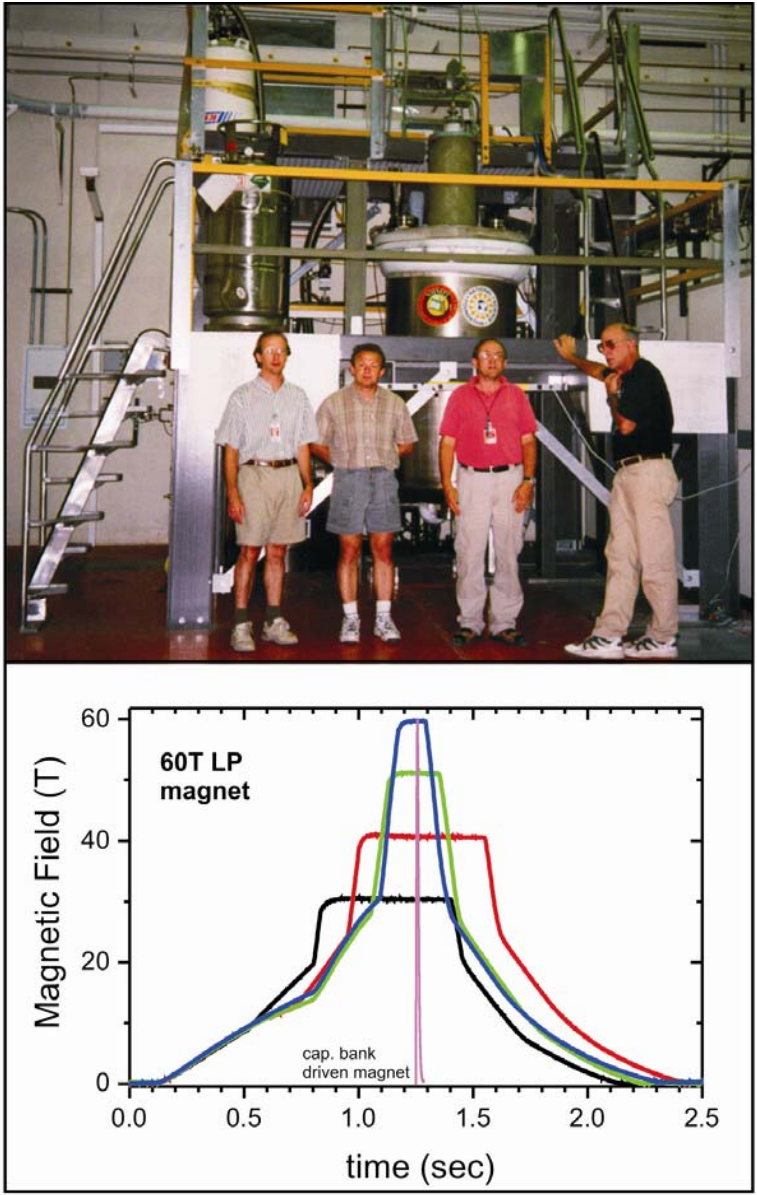

Figure 2

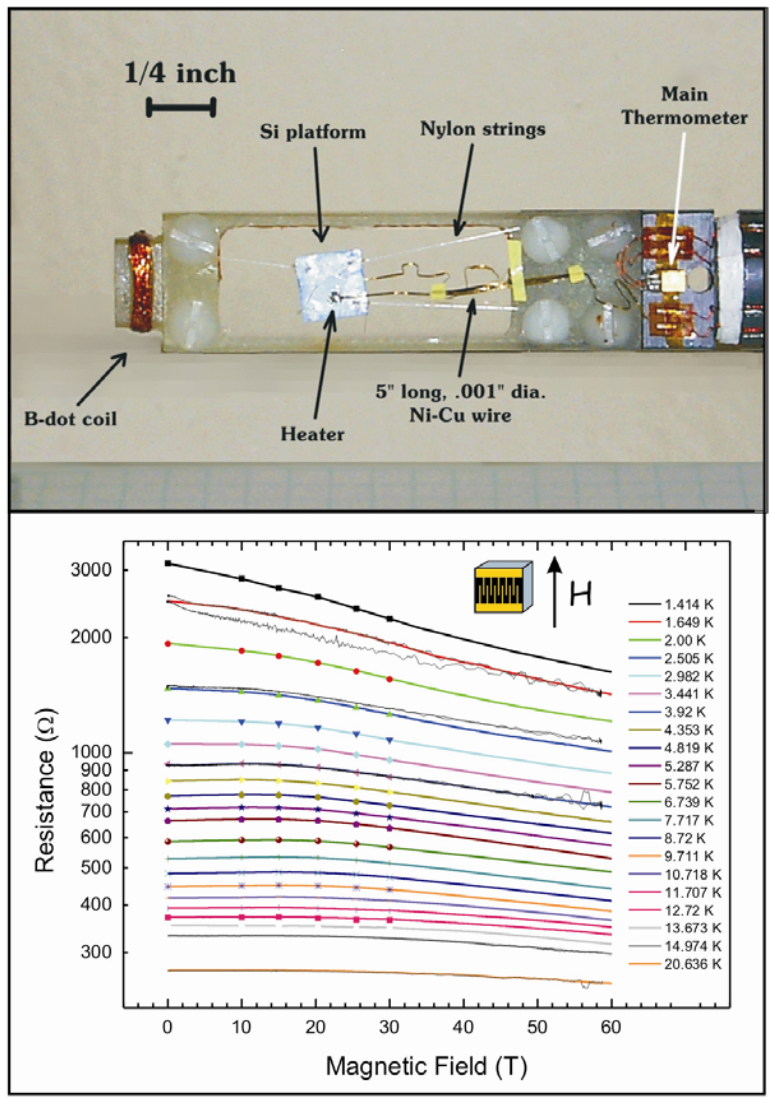

Figure 3

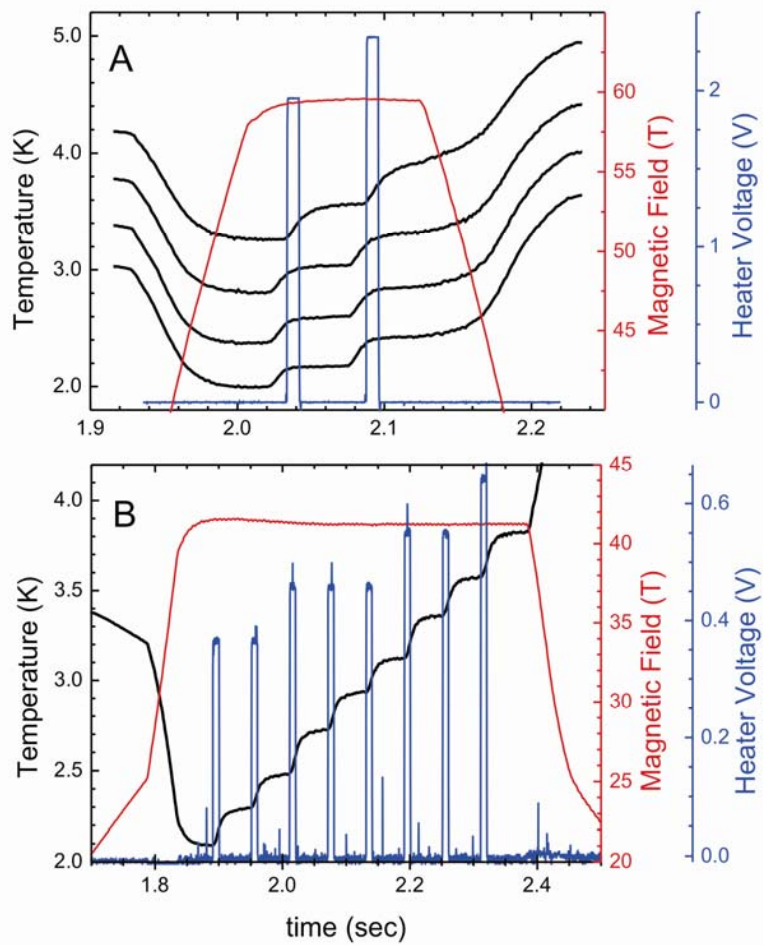

Figure 4

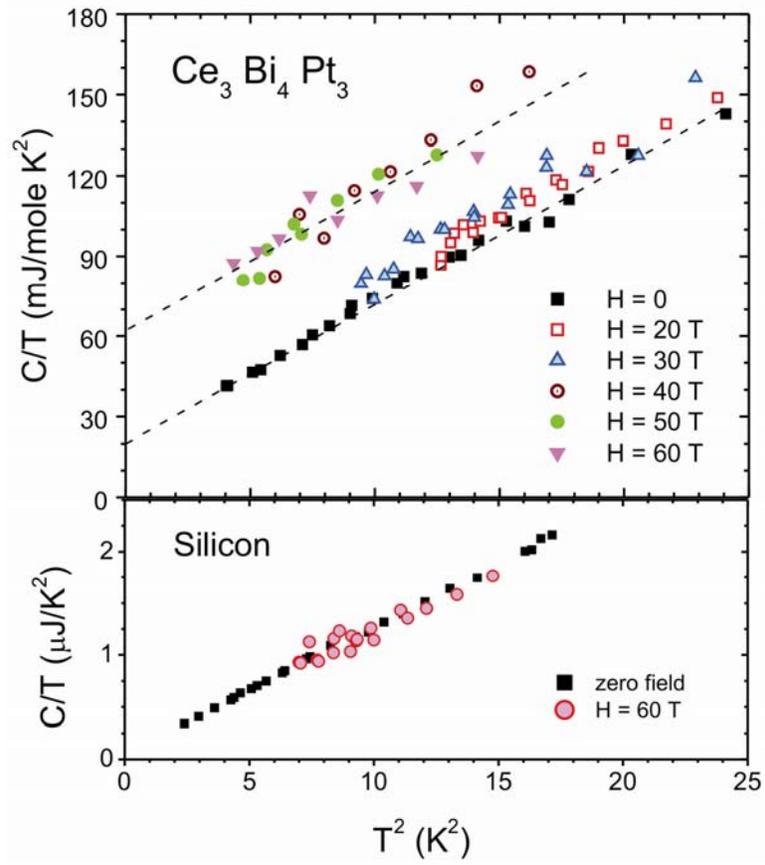

Figure 5

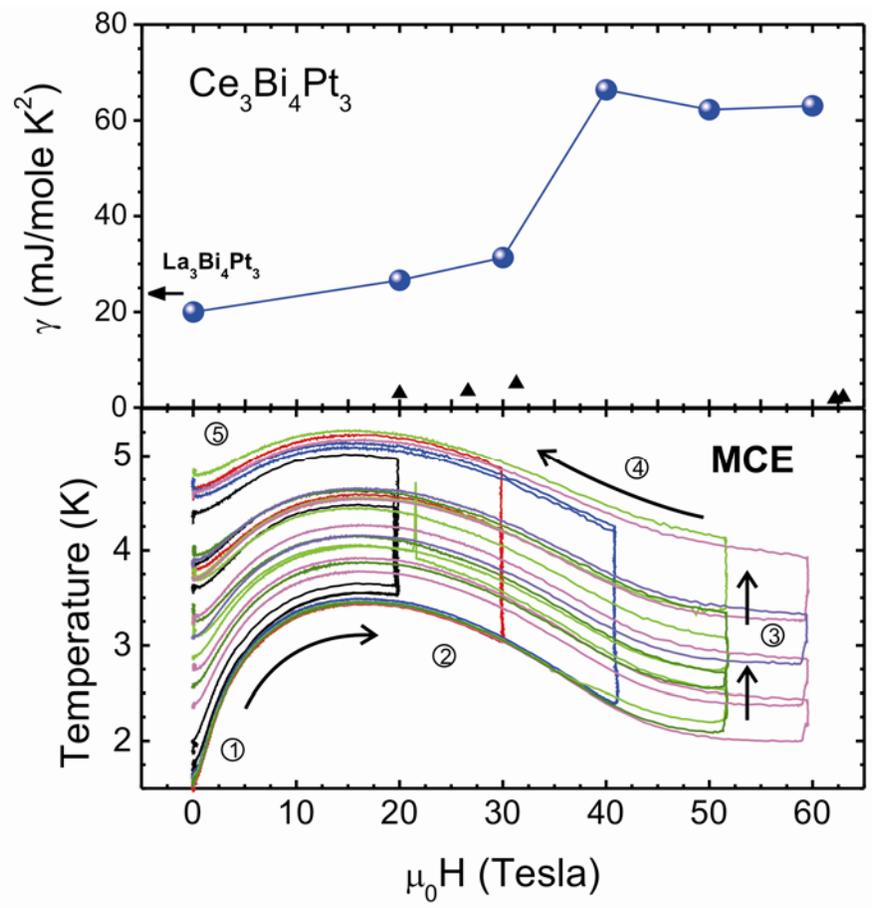

Figure 6

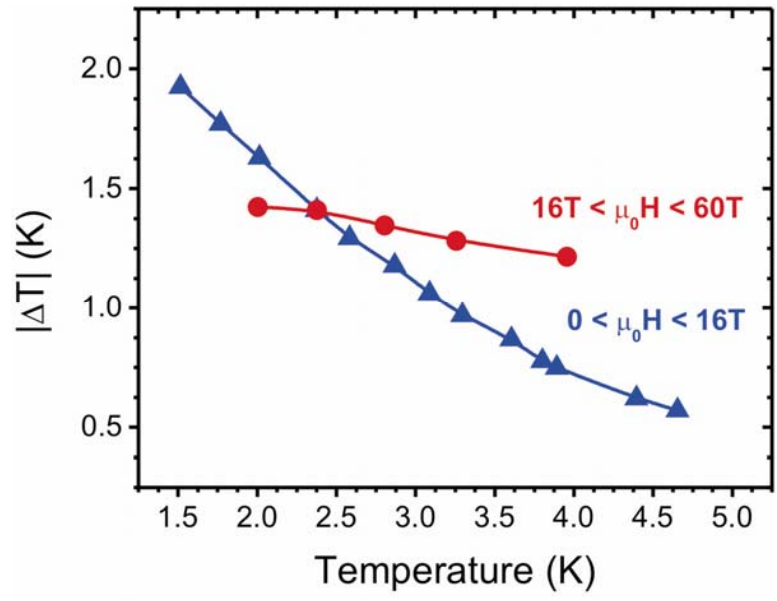

Figure 7

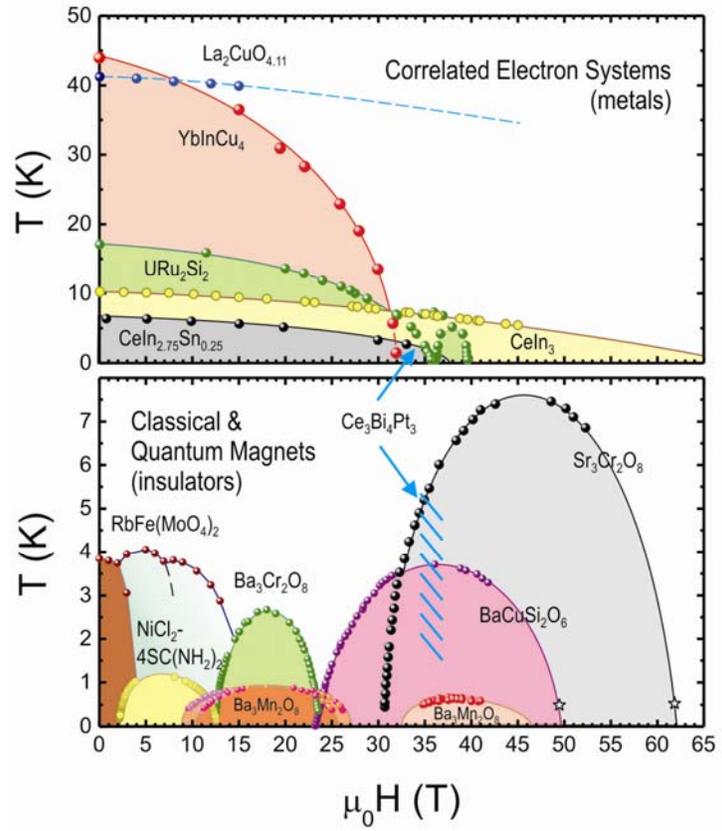

Fig 8